\documentclass[10pt,journal,compsoc]{IEEEtran}

\ifCLASSOPTIONcompsoc
  \usepackage[nocompress]{cite}
\else
  \usepackage{cite}
\fi

\usepackage{booktabs} 

\usepackage{float}
\usepackage{wrapfig}
\usepackage[tight,footnotesize,sf,SF]{subfigure}
\usepackage{enumerate}
\usepackage{todonotes}
\usepackage{soul}
\usepackage{url}
\usepackage{amsmath}
\usepackage{amsfonts}

\definecolor{green(ryb)}{rgb}{0.4, 0.69, 0.2}

\newcommand{\rev}[1]{\textcolor{black}{#1}}
\newcommand{\revnew}[1]{\textcolor{black}{#1}}

\usepackage[normalem]{ulem}

\usepackage[ruled]{algorithm2e} 

\SetAlFnt{\small}
\SetAlCapFnt{\small}
\SetAlCapNameFnt{\small}
\SetAlCapHSkip{0pt}

\newcommand{\texnet}{\text{tex}}
\newcommand{\refnet}{\text{ref}}

\newcommand{\R}{\mathbb{R}}

\makeatletter
\@tfor\next:=abcdefghijklmnopqrstuvwxyz\do{%
  \def\command@factory#1{%
    \expandafter\def\csname vec#1\endcsname{\mathbf{#1}}
  }
 \expandafter\command@factory\next
}

\@tfor\next:=ABCDEFGHIJKLMNOPQRSTUVWXYZ\do{%
  \def\command@factory#1{%
    \expandafter\def\csname mat#1\endcsname{\mathbf{#1}}
  }
 \expandafter\command@factory\next
}

\@tfor\next:=ABCDEFGHIJKLMNOPQRSTUVWXYZ\do{%
  \def\command@factory#1{%
    \expandafter\def\csname set#1\endcsname{\mathcal{#1}}
  }
 \expandafter\command@factory\next
}

\def\greekmatrices#1{%
 \@for\next:=#1\do{%
    \def\X##1;{%
     \expandafter\def\csname mat##1\endcsname{\boldsymbol{\csname##1\endcsname}}
     }
   \expandafter\X\next;
  }
}
\def\greekvectors#1{%
 \@for\next:=#1\do{%
    \def\X##1;{%
     \expandafter\def\csname vec##1\endcsname{\boldsymbol{\csname##1\endcsname}}
     }
   \expandafter\X\next;
  }
}
\greekvectors{alpha,beta,iota,gamma,lambda,nu,eta,Gamma,varsigma,Phi,Theta,delta,mu}
\greekmatrices{alpha,beta,iota,gamma,lambda,nu,eta,Gamma,varsigma,Phi,Theta,mu}

\makeatother
\begin{document}

\title{Neural Human Video  Rendering  \\ by  Learning  Dynamic  Textures  and  Rendering-to-Video  Translation}

\author{Lingjie Liu,
        Weipeng Xu,
        Marc Habermann,
        Michael Zollh\"ofer,
        Florian Bernard,
        Hyeongwoo Kim, \\
        Wenping Wang,
        Christian Theobalt
\IEEEcompsocitemizethanks{
\IEEEcompsocthanksitem This work was done when L. Liu was an intern at Max Planck Institute for Informatics. 

\IEEEcompsocthanksitem L. Liu and W. Wang are with the Department of Computer Science, The University of Hong Kong, Hong Kong, P.R.China.
E-mail: liulingjie0206@gmail.com, wenping@cs.hku.hk

\IEEEcompsocthanksitem W. Xu, M. Habermann, F. Bernard, H. Kim, C. Theobalt are with the Graphics, Vision and Video Group at Max Planck Institute for Informatics, 66123 Saarbrücken, Germany.
E-mail: wxu@mpi-inf.mpg.de, mhaberma@mpi-inf.mpg.de, fbernard@mpi-inf.mpg.de, hyeongwoo@mpi-inf.mpg.de, theobalt@mpi-inf.mpg.de

\IEEEcompsocthanksitem M. Zollh\"ofer is with the Department of Computer Science, Computer Graphics Laboratory at Stanford University, CA 94305, United States. 
E-mail: zollhoefer@cs.stanford.edu

}
}

\markboth{
}%
{Shell \MakeLowercase{\textit{et al.}}: Bare Demo of IEEEtran.cls for Computer Society Journals}

\IEEEtitleabstractindextext{%
\begin{abstract}
Synthesizing realistic videos of humans using neural networks has been a popular alternative to the conventional graphics-based rendering pipeline due to its high efficiency.
Existing works typically formulate this as an image-to-image translation problem in 2D screen space, which leads to artifacts such as over-smoothing, missing body parts, and temporal instability of fine-scale detail, such as pose-dependent wrinkles in the clothing.
In this paper, we propose a novel human video synthesis method that approaches these limiting factors by explicitly disentangling the learning of time-coherent fine-scale details from the embedding of the human in 2D screen space.
More specifically, our method relies on the combination of two convolutional neural networks (CNNs).
Given the pose information, the first CNN predicts a dynamic texture map that contains time-coherent high-frequency details, and the second CNN conditions the generation of the final video on the temporally coherent output of the first CNN.
We demonstrate several applications of our approach, such as human reenactment and novel view synthesis from monocular video, where we show significant improvement over the state of the art both qualitatively and quantitatively.
\end{abstract}

}

\maketitle

\IEEEdisplaynontitleabstractindextext

\IEEEpeerreviewmaketitle

\section{Introduction} 

\begin{figure*}
  \includegraphics[width=\linewidth]{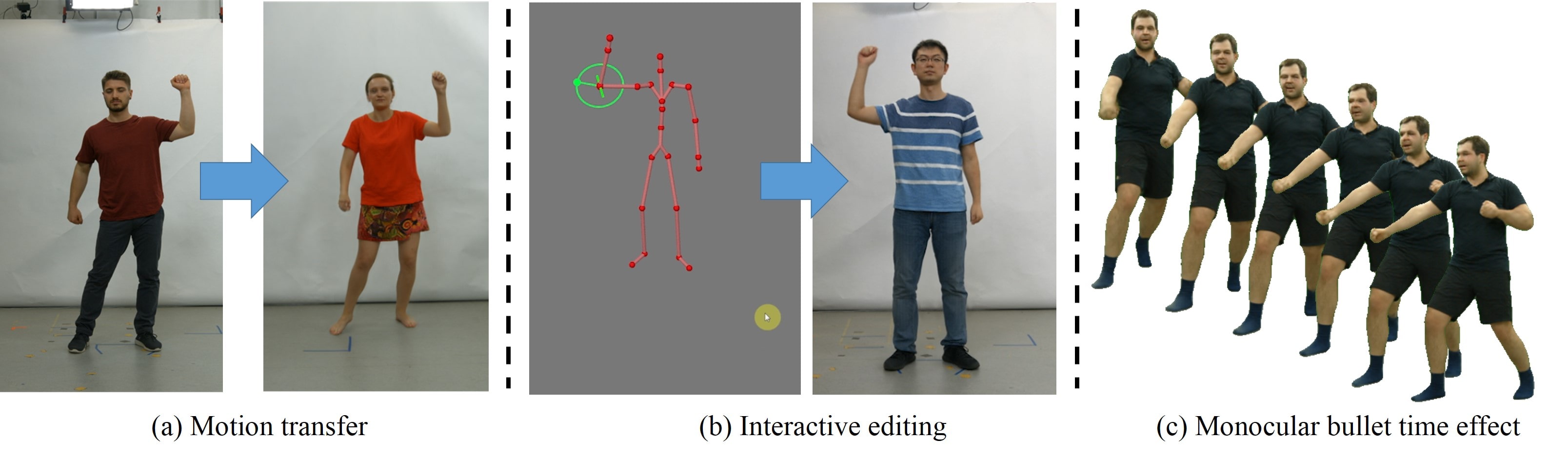}
\vspace{-0.8cm}
\caption{\rev{We present an approach for synthesizing realistic videos of humans. Our method allows for: a) motion transfer between a pair of monocular videos, b) interactively controlling the pose of a person in the video, and c) monocular bullet time effects, where we freeze time and virtually rotate the camera.}}
  \label{fig:teaser}
\end{figure*}

%
Synthesizing realistic videos of humans is an important research topic in computer graphics and computer vision, which has a broad range of applications in visual effects (VFX) and games, virtual reality (VR) and telepresence, AI assistants, and many more.
In this work, we propose a novel machine learning approach for synthesizing a realistic video of an actor that is driven from a given motion sequence.
Only a monocular video and a personalized template mesh of the actor are needed as input.
The motion of the actor in the target video can be controlled in different ways. For example by transferring the motion of a different actor in a source video, or by controlling the video footage directly based on an interactive handle-based editor.

%
Nowadays, the de-facto standard for creating video-realistic animations of humans follows the conventional graphics-based human video synthesis pipeline based on highly detailed animated 3D models.
The creation of these involves multiple non-trivial, decoupled, manual and time-consuming steps:
These include 3D shape and appearance scanning or design, hand-design or motion capture of target motions and deformations, and time-consuming photorealistic rendering.
Aiming to streamline and expedite this process, in recent years graphics and vision researchers developed data-driven methods to generate realistic images \cite{MaSJSTV2017,Balakrishnan2018,Esser2018} and videos \cite{Chan2018,Lischinski2018,wang2018vid2vid,SiaroSLS2017,Liu2018Neural} of humans.
Many of these use variants of adversarially trained convolutional neural networks to translate coarse conditioning inputs, which encode human appearance and/or pose, into photo-realistic imagery.

A prominent problem with existing methods is that fine-scale details are often over-smoothed and temporally incoherent, e.g. wrinkles often do not move coherently with the garments but look like lying on a separated spatially fixed layer floating in the screen space (see the supplementary video).
While some approaches try to address these challenges by enforcing temporal coherence in the adversarial training objective \cite{Chan2018,Lischinski2018,wang2018vid2vid}, we argue that most problems are due to a combination of two limiting factors:
1) Conditioning input is often a very coarse and sparse 2D or 3D skeleton pose rather than a more complete 3D human animation model.
2) Image translation is learned only in 2D screen space.
This fails to properly disentangle appearance effects  from residual image-space effects that are best handled by 2D image convolutions.
Since appearance effects are best described on the actual 3D body surface, they should be handled by suitable convolutions that take the manifold structure into account.
As a consequence of these effects, networks struggle to jointly generate results that show both, complete human body imagery without missing body parts or silhouette errors, as well as plausible temporally coherent high-frequency surface detail. 

%
We propose a new human video synthesis method that tackles these limiting factors and explicitly disentangles learning of time-coherent pose-dependent fine-scale detail from the time-coherent pose-dependent embedding of the human in 2D screen space.
Our approach relies on a monocular training video of the actor performing various motions, and a skinned person-specific template mesh of the actor. The latter is used to capture the shape and pose of the actor in each frame of the training video using an off-the-shelf monocular performance capture approach.
Our video synthesis algorithm uses a three-stage approach based on two CNNs and the computer graphics texturing pipeline:
1) Given the target pose in each video frame encoded as a surface normal map of the posed body template, the first CNN is trained to predict a dynamic texture map that contains the pose-dependent and time-coherent high-frequency detail.
In this normalized texture space, local details such as wrinkles always appear at the same uv-location, since the rigid and articulated body motion is already factored out by the monocular performance capture algorithm, which significantly simplifies the learning task.
This frees the network from the task of having to synthesize the body at the right screen space location, leading to temporally more coherent and detailed results.
2) We apply the dynamic texture on top of the animated human body model to render a video of the animation that exhibits temporally stable high-frequency surface details, but that lacks effects that cannot be explained by the rendered mesh alone.
3) Finally, our second CNN conditions the generation of the final video on the temporally coherent output of the first CNN. This refinement network synthesizes foreground-background interactions, such as shadows, naturally blends the foreground and background, and corrects geometrical errors due to tracking/skinning errors, which might be especially visible at the silhouettes.

To the best of our knowledge, our approach is the first dynamic-texture neural rendering approach for human bodies  that  disentangles human video synthesis into explicit texture-space and image-space neural rendering steps: pose-dependent neural texture generation and rendering to realistic video translation. This new problem formulation yields more accurate human video synthesis results, which better preserve the spatial, temporal, and geometric coherence of the actor's appearance compared to existing state-of-the-art methods.

\revnew{As shown in Figure \ref{fig:teaser}, our approach can be utilized in various applications, such as human motion transfer, interactive reenactment and novel view synthesis from monocular video. In our experiments, we demonstrate these applications and show that our approach is superior to the state of the art both qualitatively and quantitatively. }

Our main contributions are summarized as follows:
\begin{itemize}
	\item A novel three-stage approach that disentangles learning pose-dependent fine-scale details from the pose-dependent embedding of the human in 2D screen space.
	\item High-resolution video synthesis of humans with controllable target motions and temporally coherent fine-scale detail.
\end{itemize}

\section{Related work}

In the following, we discuss human performance capture, classical video-based rendering, and learning-based human performance cloning, as well as the underlying image-to-image translation approaches based on conditional generative adversarial networks.

\textbf{Classical Video-based Characters.}
Classically, the domain gap between coarse human proxy models and realistic imagery can be bridged using image-based rendering techniques.
These strategies can be used for the generation of video-based characters \cite{Xu:SIGGRPAH:2011,Li:2017,Casas:2014,Volino2014} and enable free-viewpoint video \cite{Carranza:2003,borshukov2005universal,Li:2014,zitnick2004high,collet2015high}.
Even relightable performances can be obtained \cite{LiWSLVDT13} by disentangling illumination and scene reflectance.
The synthesis of new body motions and viewpoints around an actor is possible \cite{Xu:SIGGRPAH:2011} with such techniques.

\textbf{Modeling Humans from Data.}
Humans can be modeled from data using mesh-based 3D representations.
For example, parametric models for different body parts are widely employed \cite{Blanz99,FLAME:2017,Berard14,Wood16,Wu16b,MANO:2017} in the literature.
Deep Appearance Models \cite{Lombardi:2018} learn dynamic view-dependent texture maps for the human head.
The paGAN \cite{Nagano:2018} approach builds a dynamic avatar from a single monocular image.
Recently, models of the entire human body have become popular \cite{Anguelov:2005,SMPL:2015}.
There are also some recent works on cloth modeling \cite{PonsMollSiggraph2017, Lahner_2018_ECCV, Yang_2018_ECCV}. 
One drawback of these models is that they do not model the appearance of dressed humans, i.e., the color of different garments.
To tackle this problem, generative models based on neural networks have been applied to directly synthesize 2D images of humans without having to model the 3D content.
First, these approaches have been applied to individual parts of the human body \cite{Shrivastava2017,Mueller2017,kim2018DeepVideo}.
Also models that capture the appearance of clothing have been proposed \cite{Lassner:GP:2017}.
Nowadays, similar techniques are applied for the complete human body, i.e., for the synthesis of different poses \cite{MaSJSTV2017,SiaroSLS2017,Balakrishnan2018,Esser2018}.
In contrast to previous approaches, we employ dense conditioning and learn dynamic high-frequency details in texture space to enable the temporally coherent generation of video.

\textbf{Deep Video-based Performance Cloning.}
Very recently, multiple approaches for video-based human performance cloning have been proposed \revnew{\cite{Liu2018Neural,Chan2018,Lischinski2018,wang2018vid2vid, Si_2018_CVPR, Pumarola_2018_CVPR, Esser_2018_ECCV_Workshops}} that output realistic video sequences.
These approaches learn complex image-to-image mappings, i.e., from renderings of a skeleton \revnew{\cite{Chan2018, Pumarola_2018_CVPR, Esser_2018_ECCV_Workshops, Si_2018_CVPR}}, dense mesh \cite{Liu2018Neural,wang2018vid2vid}, or joint position heatmaps \cite{Lischinski2018}, to real images.
Liu et al.~\cite{Liu2018Neural} proposed to translate simple synthetic computer graphics renderings of a human character into realistic imagery.
\textit{Everybody Dance Now} \cite{Chan2018} predicts two consecutive video frames and employs a space-time discriminator to obtain temporally more coherent synthesis results.
Deep performance cloning \cite{Lischinski2018} combines paired and unpaired training based on a two-branch network for better generalization.
The vid2vid \cite{wang2018vid2vid} approach learns high-resolution video-to-video translation based on a sequential RNN generator and uses optical flow for explicitly forward warping the last frame estimate.
All theses approaches learn an image-to-image mapping in 2D screen space based on a set of 2D convolution and deconvolution kernels.
We argue that many artifacts of these approaches, e.g., the synthesized images are over-smoothed and temporally incoherent in fine-scale detail, are due to two limiting factors: 1) Only sparse 2D or 3D skeleton conditioning and 2) learning image translation in 2D screen space. 
In contrast to existing methods, we tackle these limiting factors and explicitly disentangle learning of time-coherent pose-dependent detail in texture space from the pose-dependent embedding of the human in 2D screen space.

\textbf{Surface-based Modeling with Deep Learning.} 
Several previous works have  integrated neural synthesis into surface-based modeling~\cite{Neverova_2018_ECCV, Shysheya_2019_CVPR, guler2018densepose, Thies2019DeferredNR, li2019dense, lwb2019}. 
Deferred Neural Rendering~\cite{Thies2019DeferredNR} proposed an end-to-end training strategy to learn neural textures and deferred neural rendering jointly. They produced photo-realistic renderings for static scenes and faces with imperfect 3D reconstructed geometry. 
Some works also focus on neural synthesis for human bodies. For example, Densepose~\cite{guler2018densepose} predicts UV coordinates of image pixels from the RGB inputs, and the works \cite{Shysheya_2019_CVPR, li2019dense, lwb2019} synthesize a new image of a person in a given pose based on a single image of that person. This is done by estimating dense 3D appearance flow to guide the transfer of pixels between poses.
Textured Neural Avatars \cite{Shysheya_2019_CVPR} learns full body neural avatars with static textures based on pretrained Densepose~\cite{guler2018densepose} results. 
In contrast, our work aims at generating dynamic textures for photo-realistic renderings of human bodies, which is a more challenging task.

\textbf{3D Performance Capture of Humans. }
Monocular data based on recent performance capture techniques can provide the paired training corpora required for learning video-based performance cloning.
Historically, 3D human performance capture has been based on complex capture setups, such as multi-view reconstruction studios with a large number of cameras \cite{matusik2000image,starck2007surface,waschbusch2005scalable,cagniart2010free,vlasic2009dynamic}.
The highest quality approaches combine active and passive depth sensing \cite{collet2015high,Dou:2016,Dou:2017,wang2016capturing}.
Recent dense tracking approaches build on top of joint detections, either in 2D \cite{pishchulin2016deepcut,wei2016convolutional}, in 3D \cite{zhou2016deep,mehta2016monocular,pavlakos2016coarse}, or a combination thereof \cite{elhayek2015efficient,rosales2006combining,VNect_SIGGRAPH2017}.
The set of sparse detections provides initialization for optimization-based tracking approaches to start near the optimum to facilitate convergence.
Many approaches simplify performance capture by tracking only the degrees of freedom of a low-dimensional skeleton \cite{gall2009motion,vlasic2008articulated,liu2011markerless}, thus resolving some of the ambiguities of truly dense capture.
There is also a trend of using a reduced number of cameras, aiming to bring human performance capture to a commodity setting.
For example, some approaches enable capturing human performances from two \cite{wu2013onset} or a sparse set of cameras \cite{de2008performance}.
Recently, even lighter approaches \revnew{\cite{Zhang2014,bogo2015detailed,Helten:2013,yu2017bodyfusion, bogo2016smpl, 8491000, Pavlakos_2018_CVPR}} have been developed to deal with the rising demand for human performance capture in commodity settings, e.g., to enable virtual and augmented reality applications.
Monocular dense 3D human performance capture \cite{MonoPerfCap_SIGGRAPH2018} is still a popular research problem, with recently real-time performance being demonstrated for the first time \cite{reticam2018}.

\textbf{Conditional Generative Adversarial Networks. }
Generative adversarial networks (GANs) \cite{GoodfPMXWOCB2014,RadfoMC2016,MirzaO2014,IsolaZZE2017} have been very successful in learning to generate arbitrary imagery using a generator network based on convolutional neural networks with an encoder-decoder structure \cite{HintoS2006}. They either start from scratch using a random vector \cite{GoodfPMXWOCB2014,RadfoMC2016}, or they learn conditional image-to-image synthesis based on an input image from a different domain \cite{MirzaO2014,IsolaZZE2017}. 
U-Nets \cite{RonneFB2015} with skip connections are often employed as generator networks.
The discriminator network is trained based on a binary classification problem \cite{GoodfPMXWOCB2014} or is patch-based \cite{IsolaZZE2017}.
The generator and the discriminator are jointly trained based on a minimax optimization problem.
Very recently, high-resolution images have been generated using GANs \cite{KarraALL2018,WangLZTKC2018} with a progressive training strategy and using cascaded refinement networks \cite{ChenK2017}.
While most of these techniques are trained in a fully supervised manner based on paired training data, some approaches tackle the harder problem of learning the translation between two domains based on unpaired data \cite{ZhuPIE2017,YiZTG2017,LiuBK2017,choi2017stargan}.
Some recent works studied the problem of video-to-video synthesis. Vid2vid~\cite{wang2018vid2vid} learns high-resolution video-to-video translation based on a sequential RNN generator and uses optical flow for explicitly forward warping the last frame estimate.
The recently proposed Recycle-GAN \cite{Bansal2018} approach enables unpaired learning of a coherent video-to-video mapping.

In our work, we employ two vid2vid networks, where the first network has the task of generating a time-coherent texture with high-frequency details (e.g.~in clothing), and the second network has the task of producing the final output image by refining a rendering of a mesh that is textured with the output of the first network.
\begin{figure*}
	\includegraphics[width=\linewidth]{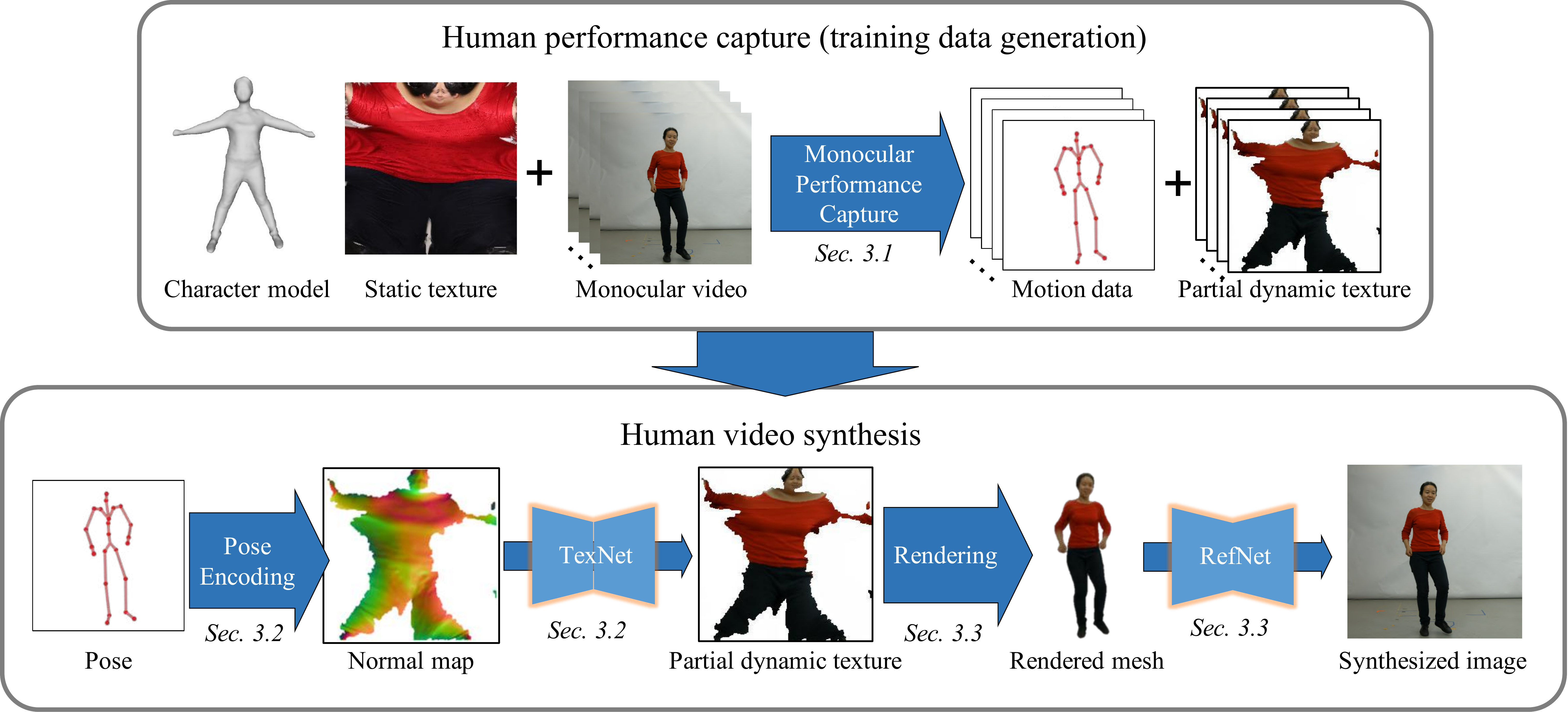}
	\caption
	{Overview of our approach. The top shows the human performance capture stage that is used for training data generation. Here, a parametric human character model is used in combination with a static texture for tracking the human motion in a monocular video \revnew{and encoding the motion to the partial normal map}. The output are motion data and dynamic (per-frame) partial textures, which capture pose-dependent high-frequency details (e.g.~cloth wrinkles).
		The bottom part shows the human video synthesis stage. 
		First, a pose-dependent partial normal map is \revnew{generated by animating the 3D static template according to the motion data and unwarping the visible region of the human body mesh to uv space} (e.g. obtained by motion capture as on the top, user-defined, or from any other source).
		This partial normal map serves as a pose encoding in texture space, which is then used as input to a \emph{texture synthesis network} (TexNet) for computing a pose-dependent partial texture map.
		The mesh rendered with this texture is then used as input to the \emph{refinement network} (RefNet) that produces the final output by blending the foreground and background, modelling shadows, and correcting geometric errors.
	}
	\label{fig:pipeline}
\end{figure*}

\section{Method} 
In this section we describe our neural human video synthesis approach.
As illustrated in Fig.~\ref{fig:pipeline}, given a monocular video of \revnew{a performing actor} and a textured mesh template of the actor, our method learns a person-specific embedding of the actor's appearance.
To generate the training data, we first employ an off-the-shelf monocular human performance capture method \cite{reticam2018} to track the  motion of the actor in the video (Sec.~\ref{sec:datamodel}).
Based on the tracking results, we generate the (partial) dynamic texture by back-projecting the video frames to the animated template mesh.
Having the motion data, partial dynamic textures, and the original video frames as the training corpus, our approach proceeds in three stages:
In the first stage, we train our \emph{texture synthesis network (TexNet)} to regress a partial texture image, which depicts the  \emph{pose-dependent} appearance details, such as wrinkles, given a certain pose as input.
Here, the pose information is encoded in a (partial) normal map in the uv-space in order to obtain an \emph{image-based pose encoding in texture space}.
In the second stage, we complete the predicted \emph{partial} texture image to a \emph{complete} texture image (Sec.~\ref{sec:dynamictexture}), and render the mesh with this \emph{complete} texture image. 
In the third stage, we translate the renderings into a realistic video with our \emph{refinement network (RefNet)} (Sec.~\ref{sec:videosynthesis}).
During testing, our method takes a motion clip from arbitrary sources (e.g., motion capture, artist-designed, etc.), and generates a video of the actor performing the input motion.

\subsection{Training Data Generation}\label{sec:datamodel}
In this section we describe the human character model, how its texture mapping is obtained, and how the human motion is captured.

\textbf{Image Sequence.}
Let $\setI_{1},\ldots, \setI_f$ be a given image sequence comprising $f$ frames of a human actor that performs motions. The $j$-th frame $\setI_j \in [0,1]^{w\times h \times 3}$ is an RGB image of dimension $w \times h$.

\textbf{3D Character Model.} 
For each subject we create a 3D character model based on the multi-view image-based 3D reconstruction software Agisoft Photoscan\footnote{\url{http://www.agisoft.com/}}.
To this end, we capture approximately a hundred images from  different view points of the actor in a static neutral pose (upright standing and the arms forming a ``T-pose'', see Fig.~\ref{fig:pipeline} ``Character model'').
This data is then directly used as input to Photoscan, which produces a textured 3D model of the person, as shown in Fig.~\ref{fig:pipeline} (``Character model'' and ``Static texture'').
Then, we rig the character model with a parameterized skeleton model, similarly as done in other approaches (e.g.~ \cite{reticam2018}).
Based on this procedure we obtain a parameterized surface mesh model with vertex positions $\setM(\theta) \in \mathbb{R}^{n \times 3}$, where $n$ is the number of mesh vertices and $\theta \in \R^{33}$ is the pose parameter vector, where among the $33$ scalar values  $6$ are global rigid pose parameters, and $27$ are pose articulation parameters in terms of joint angles.

\textbf{Texture Mapping.}
For texture mapping, we unwrap the human body surface mesh and map it onto the unit square $[0,1]^2$ using the quasi-harmonic surface parameterization method of~\cite{zayer2005discrete}, which reduces the parametric distortion by attempting to undo the area distortion in the initial conformal mapping.
To this end, the mesh is first cut along the spine, followed by two cuts along the legs, as well as three cuts along the arms and the head.
Then, this boundary is mapped to the boundary of the square.
A so-created RGB texture $\setT \in  [0,1]^{w \times h \times 3}$ is shown in Fig.~\ref{fig:pipeline} (``Static texture'').

\textbf{Human Performance Capture.}
We employ the recent real-time dense motion capture method of ~\cite{reticam2018}.
Their two-stage energy-based method first estimates the actor's pose by using a sparse set of body and face landmarks, as well as the foreground silhouette.
The output of the motion capture stage is the pose vector $\theta$, which can be used to pose the surface model, resulting in a deformed mesh with vertex positions $\setM(\theta)$.
Next, the reconstruction is refined on the surface level to account for local non-rigid deformations that cannot be captured by a pure skeleton-based deformation.
To this end, per-vertex displacements are estimated using a dense silhouette and photometric constraints.

\textbf{Target Dynamic Texture Extraction.} 
After the performance capture, we generate the pose-specific partial dynamic texture $\setT_j$ by back-projecting the input image frame $\setI_j$ onto the performance capture result, i.e., the deformed mesh $\setM(\theta_j)$.
Note that the generated dynamic textures are incomplete, since we only have front view observation, due to the monocular setup.

\begin{figure} 
	\centerline{ 
		\subfigure{\includegraphics[width=\linewidth]{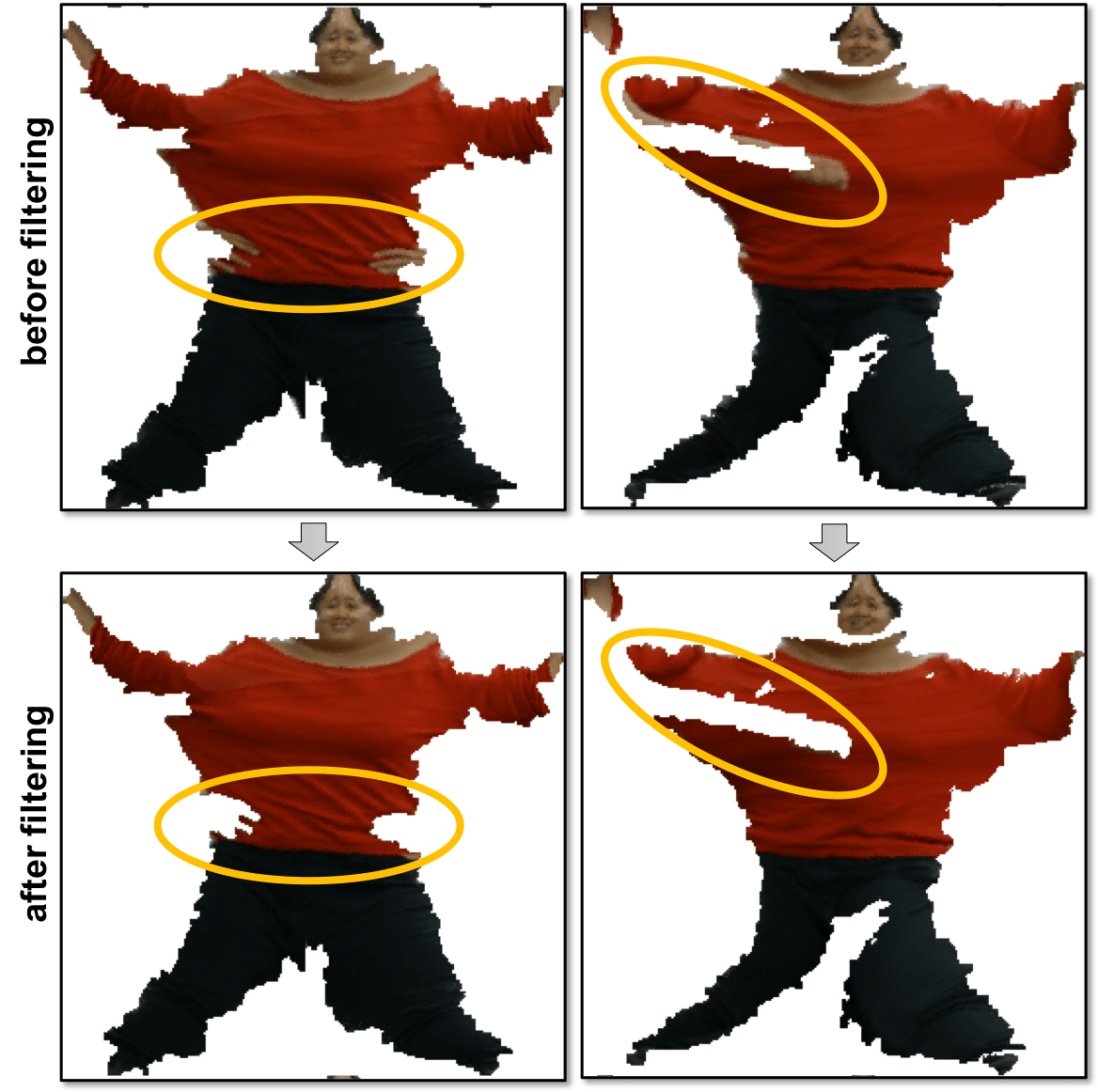} }
	}
	\caption{Effect of our filtering procedure. The top row shows the texture map before filtering, and the bottom row shows it after filtering. 
	} 
	\label{fig:filtering}
\end{figure}

Although the reconstructed 3D body model yields a faithful representation of the true body geometry, small tracking errors between the digital model and the real human are inevitable.
A major issue is that such small misalignments would directly result in an erroneous texture map $\setT_j$ (e.g. a common case is that a hand in front of the torso leads to the incorrect assignment of the hand color to a torso vertex, see Fig.~\ref{fig:filtering}).
Using such noisy texture maps  would be disadvantageous for learning, as the network would need to spend capacity on understanding and (implicitly) fixing these mismatches.
Instead, based on a simple image-based analysis we filter out the erroneous parts and thereby avoid training data corruption.
The filtering method consists of four simple steps: 
\begin{enumerate}[(i)]
	\item First, we generate an average texture map $\overline{\setT} \in [0,1]^{w \times h \times 3}$ by averaging all colors of $\setT_1,\ldots,\setT_f$ along the temporal axis. Note that texels that correspond to occluded mesh vertices of $\setM(\theta_j)$, i.e. zero values in the texture map $\setT_j$, are not taken into account for averaging.
	\item  Next we use a $k$-means clustering procedure to cluster all the colors present in the average texture map $\overline{\setT}$ so that we obtain a small number of $k$ \emph{prototype colors} that are ``typical'' to the specific sequence at hand.
	\item Then, for all frames $j$ we assign to each texel of $\setT_j$ its nearest prototype color, which is then used to compute a (per-texel) histogram of the prototype colors over all the frames (again, only considering visible parts).
	\item Finally, for each texel we check whether there is a prototype color that only occurs very rarely. If yes, we suppose that it is caused by the transient color of a tracking error (e.g.~a wrongly tracked hand), and therefore discard the color assignment for this texel in all frames where the insignificant color is present. 
\end{enumerate}
By doing so,  erroneous color assignments are excluded from the partial textures to enhance network training quality. In Fig.~\ref{fig:filtering} we illustrate the effect of our filtering procedure.
In addition, to avoid the background pixels from being projected onto the mesh, we apply a foreground mask, generated with the video segmentation method of~\cite{Cae+17}, on the input images when doing the back-projection.

\rev{Subsequently, we fill in the discarded texels based on the average texture $\overline{\setT}$.}
The so-created partial dynamic textures $\setT_j$, together with the tracking results $\theta_j$, are then used as training data to our networks.

\subsection{Dynamic Texture Synthesis}\label{sec:dynamictexture}
Now we describe our first network, the \emph{texture synthesis network} (TexNet),  which generates a \emph{pose-dependent texture} given the corresponding pose $\theta$ as conditional input.
With that, we are able to generate pose-dependent high-frequency details directly in texture space, such as for example cloth wrinkles, which otherwise would require complex and computationally expensive offline rendering approaches (e.g. cloth simulation).

\textbf{Pose Encoding.}
Since the texture that we aim to generate is represented in texture space (or \emph{uv-space}), it is advantageous to also use an input that lives in the same domain.
Hence, we have chosen to represent the pose using a \emph{partial normal map in texture space} (cf.~Fig.~\ref{fig:pipeline}, ``Partial normal map''), which we denote by $\setN \in (\setS^2)^{w\times h \times 3}$, for $\setS^2$ being a unit 2-sphere embedded in 3D space (i.e. the set of all unit length 3D vector). 
We note that here we use the camera coordinate system for normal calculation, since the appearance/illumination would change if the person faces a different direction.
In order to allow for texels that do not have an assigned normal, we include the zero vector in $\setS^2$.
Compared to other pose representations, such as for example \rev{a depth map of a 3D skeleton}, using such an \emph{image-based} pose encoding in texture space facilitates simplified learning because the network does not need to additionally learn the translation between different domains (\rev{see the ablation study}). 
The partial normal map $\setN_j$ is created based on the 3D body reconstruction $\setM(\theta_j)$ at frame $j$.

To this end, for each vertex of the fitted 3D model that is visible in the current frame, we compute its (world-space) surface normal, and then  create the partial normal map using the mesh's uv-mapping (Sec.~\ref{sec:datamodel}).
Note that those areas in the partial normal map that correspond to invisible vertices are set to zero, cf.~Fig.~\ref{fig:pipeline} (``Partial normal map'').

\textbf{Texture Synthesis Network.}
The TexNet has the purpose of creating a pose-dependent texture from a given input partial normal map, as illustrated in Fig.~\ref{fig:pipeline}.
As such, we aim to learn the network parameters $\Theta$ that parameterize the TexNet $f^\texnet_{\Theta}$ translating a given partial normal map $\setN \in (\setS^2)^{w \times h}$ to a pose-dependent texture $\setT \in [0,1]^{w \times h \times 3}$.
For training the network, we require pairs of partial normal maps and target partial texture maps $\{(\setN_j,\setT_j):1\leq j \leq f\}$,  which are directly computed from the input sequence $\setI_1,\ldots,\setI_f$ based on motion capture as described in Sec.~\ref{sec:datamodel}.
During test time, for each frame $\setI_j$ the partial normal map $\setN_j$ is extracted using the 3D reconstruction $\setM(\theta_j)$, and the texture map $\setT_j = f^\texnet_{\Theta}(\setN_j)$ is synthesized by the network.

\textbf{Network Architecture.}
Since the recent \emph{vid2vid} network~\cite{wang2018vid2vid} was shown to synthesize photo-realistic and temporally consistent videos, we build our network upon its state-of-the-art architecture.
It considers the temporal consistency in a local window (we set the window size to 3 in our experiments). This is achieved by leveraging optical flow based warping together with conditional generative adversarial networks (cGANs). The cGANs jointly learn the generator function $f^{\texnet}_{\Theta}$ to  produce the output texture map $\setT = f^{\texnet}_{\Theta}(\setN)$ from a given conditioning input partial normal map $\setN$, along with a discriminator function $\setD$. The latter has the purpose to classify whether a given texture map $\setT$ is a synthesized texture (produced by the generator $f^{\texnet}_{\Theta}$) or a real texture.
The general cGAN loss function reads: 
\begin{align} %
\label{eq:cGAN}
     \setL^{\text{cGAN}}(f^{\texnet}_{\Theta}, \setD) &= \mathbb{E}_{\setT,\setN}(\log \setD(\setT,\setN))  \\ 
     &\qquad  + \mathbb{E}_{\setN}(\log(1{-}\setD(f^{\texnet}_{\Theta}(\setN),\setN))) \,. \nonumber
\end{align}
To obtain realistic individual frames, as well as a temporally consistent sequence of frames, a per-frame cGAN loss term $\setL^{\text{frm}}$ is used in combination with a video cGAN loss term $\setL^{\text{vid}}$ that additionally incorporates the previous two frames.
Furthermore, the term $\setL^{\text{flow}}$ is used to learn the optical flow fields.
The total learning problem now reads:
\begin{align}
\label{eq:vid2vid}
    \min_{f^{\texnet}_{\Theta}} \max_{\setD^{\text{frm}},\setD^{\text{vid}}} & \setL^{\text{frm}}(f^{\texnet}_{\Theta}, \setD^{\text{frm}}) +  \setL^{\text{vid}}(f^{\texnet}_{\Theta}, \setD^{\text{vid}}) + \lambda \setL^{\text{flow}} \,.
\end{align}
\textbf{Training.}
We use approximately 12,000 training pairs, each of which consists of the ground truth texture map $\setT$ as well as the partial normal map $\setN$.
For training, we set a hyper-parameter of $\lambda=10$ for the loss function, and use the Adam optimizer ($lr = 0.0002$, $\beta_1 = 0.5$, $\beta_2 = 0.99$), which we run for a total number of 10 epochs with a batch size of 8.
For each sequence of $256 \times 256$ images, we use 8 Nvidia Tesla V100 GPUs to train for about 2 days. 

\revnew{\textbf{Runtime During Testing.}
A forward pass of TexNet takes ~8ms/frame to generate a $256\times256$ image on a Nvidia Tesla V100 GPU. }

\subsection{High-fidelity Video Synthesis}\label{sec:videosynthesis} 
By synthesizing the texture using TexNet, we bake pose-specific high-frequency details into the texture.
This texture is now used for generating the final output by means of a \emph{refinement network} (RefNet).
The RefNet has the task of synthesizing the background, as well as dealing with background-foreground interactions, such as shadows.
Moreover, it implicitly learns to correct geometric errors due to tracking misalignments and due to skinning errors.

\textbf{Training Data.}
In order to train the RefNet, we first run TexNet in order to obtain the (partial) dynamic texture map of all frames.
Subsequently, we fill in the invisible texels based on the average texture (across the temporal axis) to obtain a full texture map.
Then, we use the full texture map to render the mesh of the 3D reconstruction obtained by motion capture.
The RefNet is now trained on this data for the task of synthesizing the original input image, given the rendered mesh, cf.~Fig.~\ref{fig:pipeline}.

\textbf{Network Architecture.}
The architecture is the same as the TexNet, with the main difference being that instead of learning a function that maps a partial normal map to a color texture, we now learn a function $f_{\Phi}^{\refnet}$ that maps a rendered image to a realistic output, see Fig.~\ref{fig:pipeline}.
The loss function is analogous to Eq.~\ref{eq:vid2vid} with $f_{\Phi}^{\refnet}$ in place of $f_{\Theta}^{\texnet}$.

\textbf{Training.}
We use approximately 12,000 training pairs, each of which consists of the rendered image and the original RGB image.
For training, we set a hyper-parameter of $\lambda=10$ for the loss function, and use the Adam optimizer ($lr = 0.0002$, $\beta_1 = 0.5$, $\beta_2 = 0.99$) which we run for a total of 10 epochs with a batch size of 8.
For each sequence of $256\times256$ images, we use 8 Nvidia Tesla V100 GPUs to train for about 2 days.
For higher resolution results of $512\times 512$, we need about 6 days on the same GPUs.

\revnew{\textbf{Runtime During Testing.}
A forward pass of RefNet requires ~8ms/frame to generate $256\times 256$ images on a Nvidia Tesla V100 GPU, ~15ms/frame for $512\times 512$, and ~33ms/frame for $1024\times 1024$.}

\section{Experiments} \label{sec:experiments}

\newcommand{\methodlabel}[1]{\parbox{2.79cm}{\centering {#1}}}
\newcommand{\figCompScale}{0.33}
\begin{figure*}
\rotatebox[origin=l]{90}{ 
\methodlabel{\textbf{Ours}}
\methodlabel{\textbf{Rendering}}
\methodlabel{\textbf{Driving Motion}}
}  
  \subfigure{\includegraphics[scale=\figCompScale]{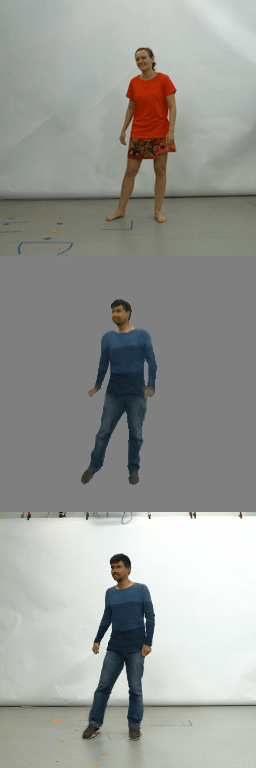}}%
  \subfigure{\includegraphics[scale=\figCompScale]{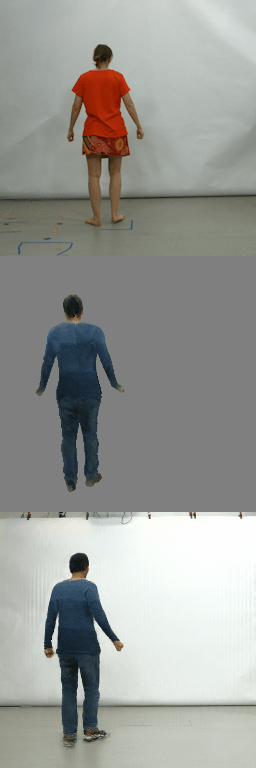}}%
  \subfigure{\includegraphics[scale=\figCompScale]{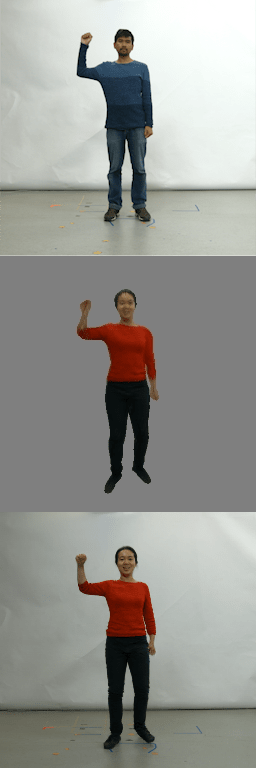}}%
  \subfigure{\includegraphics[scale=\figCompScale]{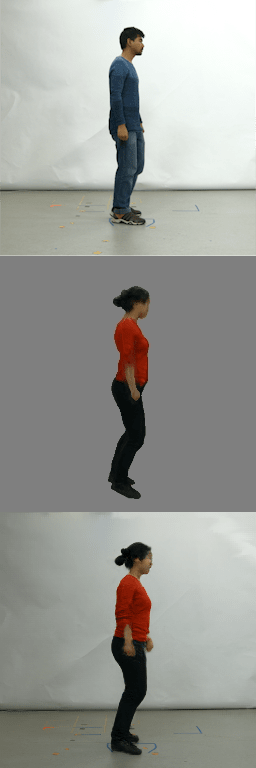}}%
  \subfigure{\includegraphics[scale=\figCompScale]{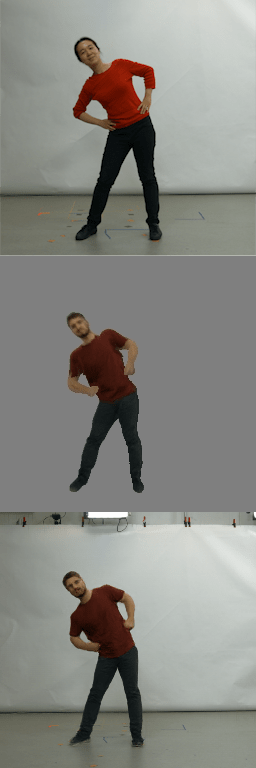}}%
  \subfigure{\includegraphics[scale=\figCompScale]{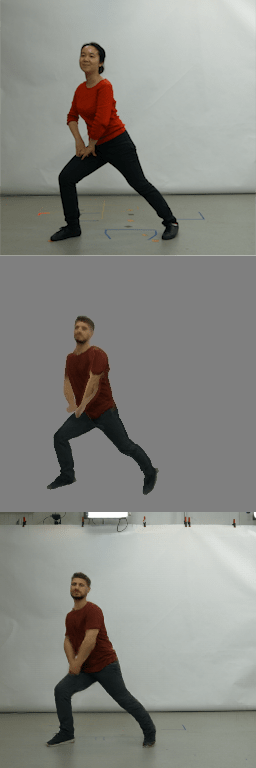}}%
\vspace{-0.3cm}
\rotatebox[origin=l]{90}{ 
\methodlabel{\textbf{Ours}}
\methodlabel{\textbf{Rendering}}
\methodlabel{\textbf{Driving Motion}}
}  
  \subfigure{\includegraphics[scale=\figCompScale]{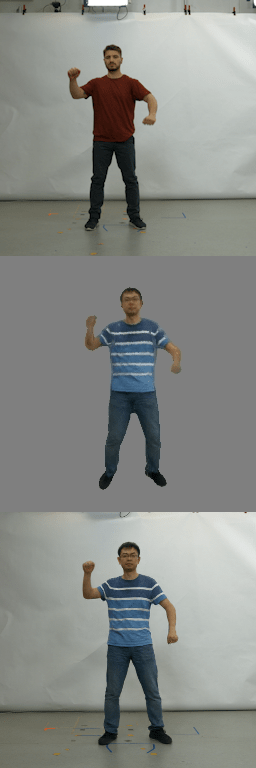}}%
  \subfigure{\includegraphics[scale=\figCompScale]{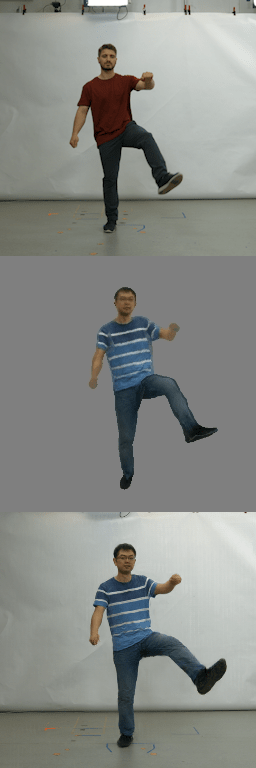}}%
  \subfigure{\includegraphics[scale=\figCompScale]{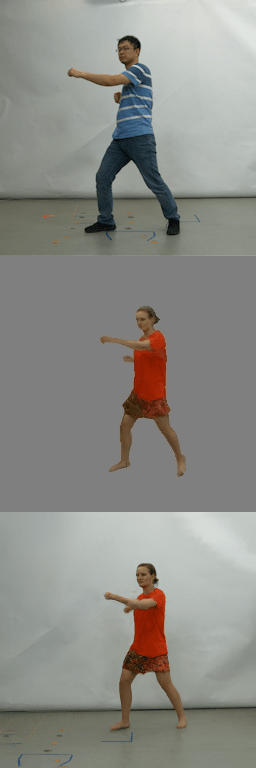}}%
  \subfigure{\includegraphics[scale=\figCompScale]{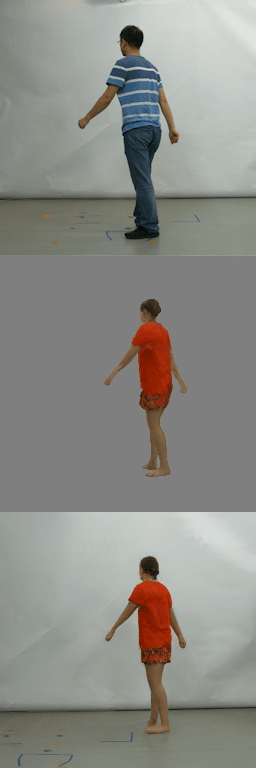}}%
  \subfigure{\includegraphics[scale=\figCompScale]{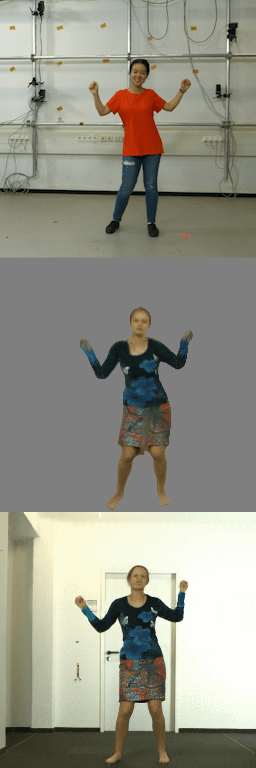}}%
  \subfigure{\includegraphics[scale=\figCompScale]{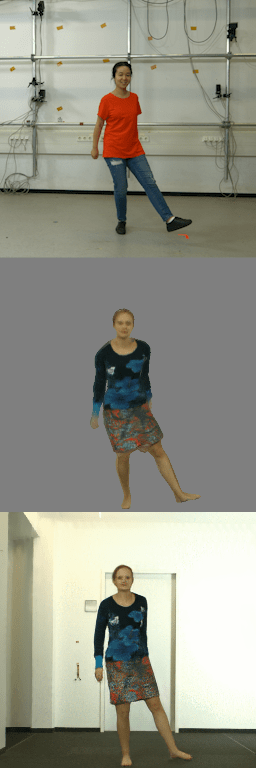}}%
  \caption{Example frames for our motion transfer results. The 1st row shows frames from the source videos, 
  the 2nd row shows the meshes rendered with the synthesized textures (input to our RefNet), and the 3rd row shows our final results. See our supplementary video for complete results. }
  \label{fig:motion_transfer_qualitative}
\end{figure*}

\begin{figure*}
\rotatebox[origin=l]{90}{ 
\methodlabel{Esser et al. \cite{Esser2018}}
\methodlabel{Ma et al. \cite{MaSGVSF2017}}
\methodlabel{Chan et al. \cite{Chan2018}}
\methodlabel{Wang et al. \cite{wang2018vid2vid}}
\methodlabel{Liu et al. \cite{Liu2018Neural}}
\methodlabel{\textbf{Ours}}
\methodlabel{\textbf{Driving Motion}}
}  
   \subfigure{\includegraphics[scale=\figCompScale]{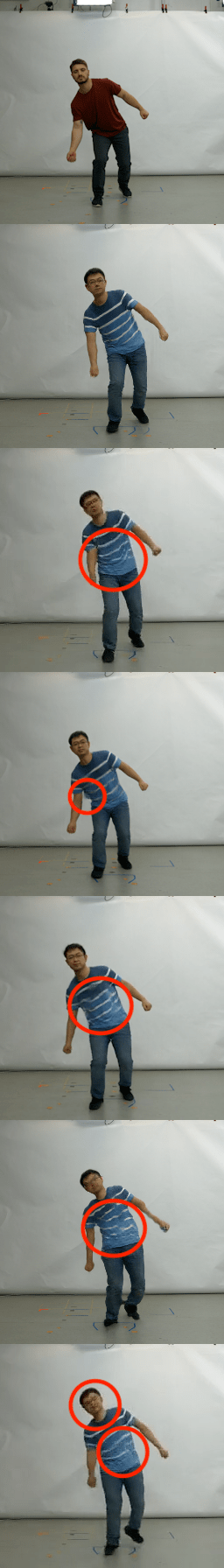}}%
   \subfigure{\includegraphics[scale=\figCompScale]{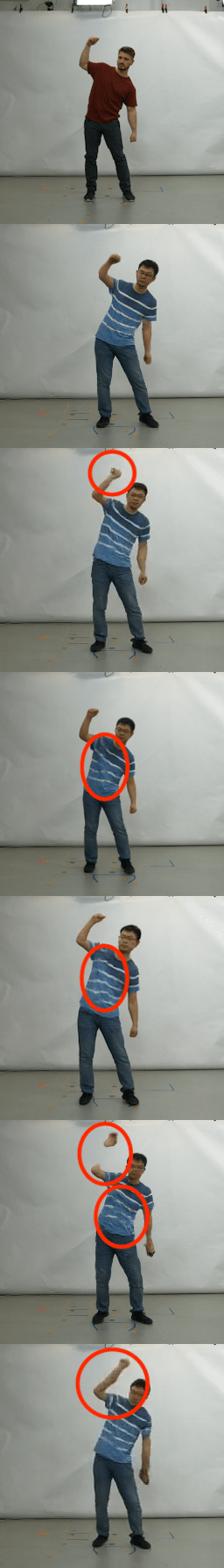}}%
   \subfigure{\includegraphics[scale=\figCompScale]{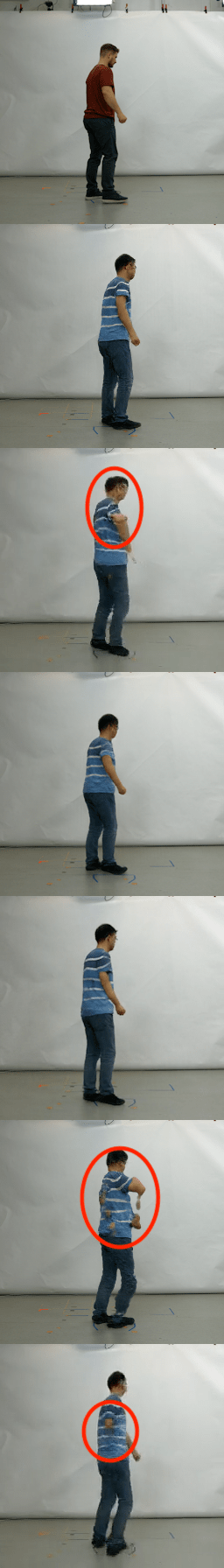}}
   \subfigure{\includegraphics[scale=\figCompScale]{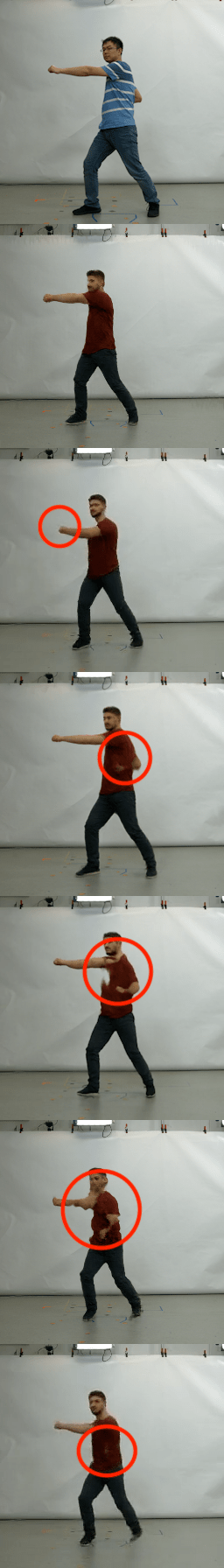}}%
   \subfigure{\includegraphics[scale=\figCompScale]{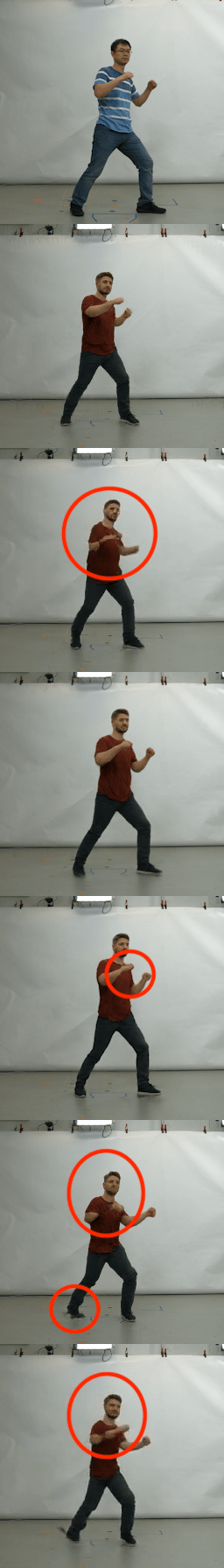}}%
   \subfigure{\includegraphics[scale=\figCompScale]{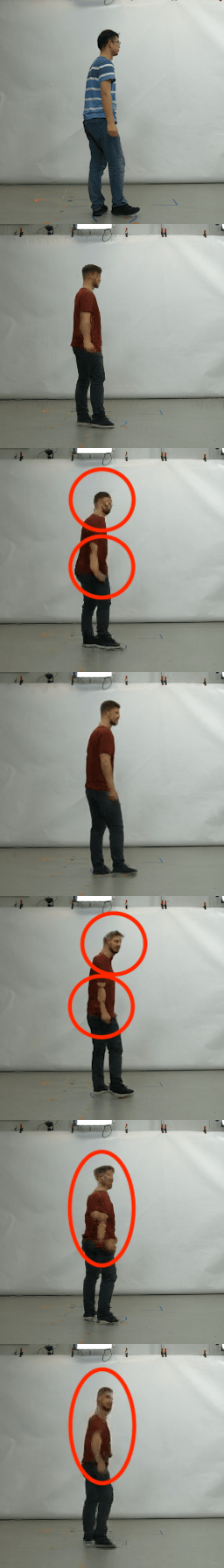}}%
    \caption{Qualitative comparison against previous state-of-the-arts on the motion transfer application. The first row shows the input sequence that is used to drive the motion, the second row shows the results obtained from our method, and the remaining rows show results obtained by the methods from~Liu et al. \cite{Liu2018Neural},~Wang et al. \cite{wang2018vid2vid},~Chan et al. \cite{Chan2018}, ~Ma et al. \cite{MaSGVSF2017},~Esser et al. \cite{Esser2018}.}
    \label{fig:motion_transfer_comparison}
\end{figure*}

To evaluate our approach and provide comparisons to existing methods, we conduct experiments on the 7 video sequences from~\cite{Liu2018Neural}.
Each sequence comprises approximately $12{,}000$ frames, where the subjects are instructed to perform a wide range of different motions, so that the space of motions is sufficiently covered by the training data.
We split each sequence into a training sequence and a test sequence, where the last quarter of each sequence is used for testing.
In addition, we captured a new sequence to demonstrate the use of our approach in a novel-view synthesis setting, and we also evaluate our method based on a community video as driving sequence.
%
%
In the following, we show our qualitative results on the motion transfer and novel-view synthesis tasks and provide comparisons to previous state-of-the-art methods.
Then, we perform an ablation study to evaluate the importance of each component of our approach.

\begin{figure*}[ht]
\includegraphics[width=\linewidth]{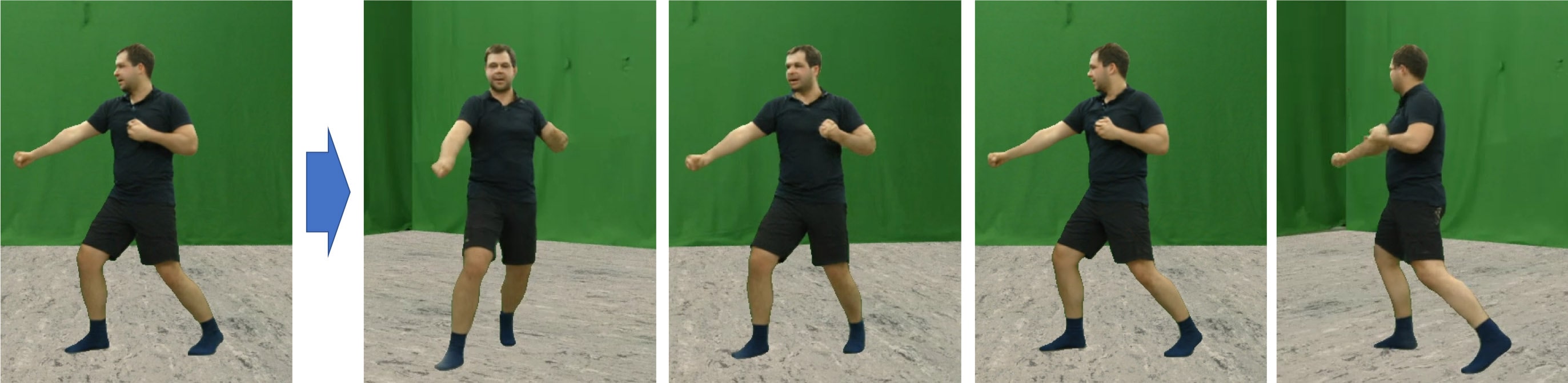}
     \caption{Bullet time video frame examples. Our method can be used to synthesize new views of the actor using just a monocular video.}
\label{fig:bullet_time}
\end{figure*}

\subsection{Motion Transfer}
\begin{figure}%
\includegraphics[width=1\linewidth]{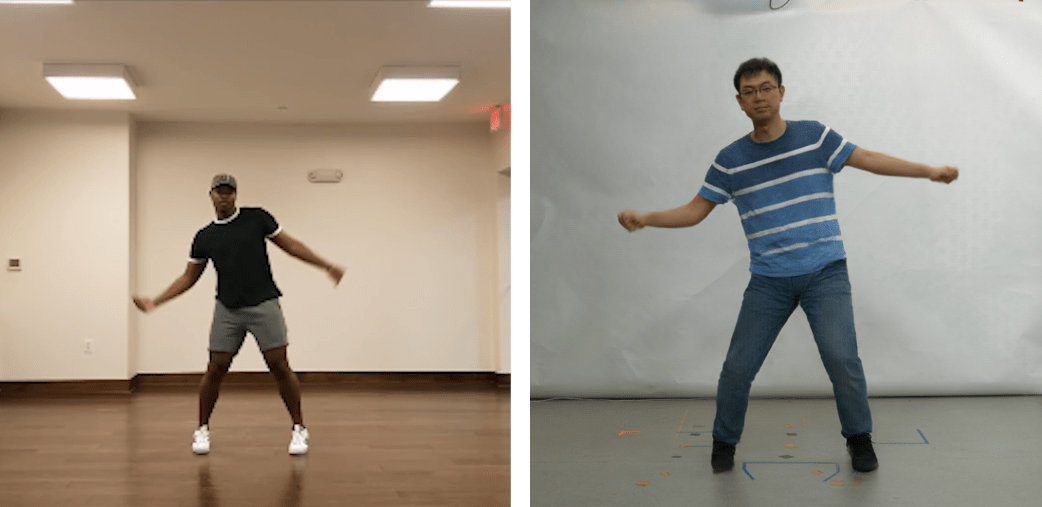}
  \caption{\rev{Reenactment result of using internet video footage as driving motion.}}
  \label{fig:dance}
\end{figure}

For the motion transfer application, we make use of pairs of monocular video sequences and our goal is to synthesize a video of the target actor performing the motion of the source actor, i.e., to transfer the motion from the source video to the target video.
To this end, we estimate the optimal pose of the target person for each frame by solving a inverse kinematics (IK) problem as in~\cite{Liu2018Neural}, which encourages the corresponding keypoints on both skeletons, including the joints and facial landmarks, to match each other in 3D as much as possible. Note that directly applying the source's skeletal pose parameters to the target skeleton may fail to produce acceptable results in general for two reasons: First, this would require that both skeletons have exactly the same structure, which may be overly restrictive in practice. Second, even more importantly, differences in the rigging of the skeleton would lead to incorrect poses if the pose parameters of the source skeleton are applied directly to the target skeleton.
Several example frames of the motion transfer results are shown in Fig.~\ref{fig:motion_transfer_qualitative}.
We can see from the mesh rendered with the synthesized dynamic texture (see Fig.~\ref{fig:motion_transfer_qualitative}, 2nd row) that our TexNet is able to capture the pose-dependant details, such as wrinkles, while the RefNet yields realistic images, where  artifacts due to tracking/skinning errors are corrected and the natural blending and interaction (shadows) between foreground and background are synthesized. \rev{We point out that even the results of non-frontal motions look plausible. }
In our supplementary video we show additional animated results.
Our approach can also take a user-designed motion as source motion input, which allows the user to interactively reenact the actor using a handle-based editor (see the demonstration in our supplementary video).
\rev{Furthermore, we stress test our approach by using internet video footage as driving motion. Although the driving motion is very different from the motions in our training corpus, our approach generates plausible results (see Fig.~\ref{fig:dance} and the supplementary video).}

We compare our approach with the following five methods on two sequences: Esser et al. \cite{Esser2018}, Ma et al. \cite{MaSGVSF2017}, Liu et al. \cite{Liu2018Neural}, Chan et al. \cite{Chan2018}, and Wang et al. \cite{wang2018vid2vid}.
\revnew{For fair comparison, we apply the input in the same formats to the networks in the comparison methods as they require. Specifically, the input to Esser et al. \cite{Esser2018}, Ma et al. \cite{MaSGVSF2017} and Chan et al. \cite{Chan2018} is the motion of a 2D skeleton. A part-based RGBD representation is used as input for Liu et al. \cite{Liu2018Neural}. The tracking results obtained with OpenPose \cite{cao2018openpose} and DensePose \cite{guler2018densepose} are used as input to Wang et al. \cite{wang2018vid2vid}. }

The qualitative comparisons are provided in Fig.~\ref{fig:motion_transfer_comparison}.
Again, we refer the reader to the supplementary video for better visual comparisons.
As can be seen from the video, our approach yields temporally more coherent results and exhibits less artifacts than the competing methods.
Especially, the artifact of missing limbs is significantly alleviated in our results.
Also note that, in contrast to our method, the methods of Esser et al. \cite{Esser2018}, Ma et al. \cite{MaSGVSF2017} and Wang et al. \cite{wang2018vid2vid} do not preserve the identity (body shape) of the actors, since their motion transfer is done in the 2D image space (e.g. with 2D landmarks positions), while ours is done in the skeleton pose space.
\revnew{Furthermore, our approach yields geometrically more consistent results. For example, wrinkles in our results move coherently with the garments, rather than being attached to a separated spatially fixed layer in screen space, as can be observed for the other methods. These benefits come from a well-designed three-stage pipeline that first generates a dynamic texture with time-coherent high-frequency details and then renders the mesh with the dynamic texture, which is eventually refined in screen space. To help understanding how each component of the pipeline contributes to the final result, we provide thorough ablations in Section \ref{ablation}, including the use of rendered mesh with dynamic texture rather than a sparse skeleton or rendered meshes with average/static texture as input to the second network, and the importance of a partial normal map as input to the first network, etc.}

\rev{We also compare the output of TexNet with the texture map retrieved by a simple nearest-neighbor-based approach. The similarity of two motions is defined as the $\ell_2$-norm of the difference of the motions represented by 30 joint angles $\theta$ ($\theta \in (-\pi, \pi]$). We fetch the texels from the texture map of the closest pose and fill-in the invisible region using the average texture. The results are clearly worse and show many spatial and temporal artifacts (see the supplementary video).}

\subsection{Novel-View Synthesis}

Novel-view synthesis is an important task for many real-world applications, such as VR-based telepresence and the iconic ``bullet time'' visual effect for the film industry.
Our proposed approach can deliver such results based on just a monocular video.
To demonstrate this, we captured a monocular video sequence and showcase the bullet time visual effect based on our approach.
In each video, the actor is asked to perform a similar set of motions (Karate exercise) for multiple repetitions in \revnew{eight} different global rotation angles \revnew{(rotated in 45 degrees steps)} with respect to the camera.
This lets the camera capture similar poses from different viewing directions.
The captured video is tracked and used for training our networks.
For testing, we select a fixed pose out of the motion sequences, and then use a virtual camera orbiting around the actor to generate the conditional input images to our approach.
This allows us to synthesize realistic video of the actor frozen in a certain pose, viewed from different angles.
Some example frames are shown in Fig.~\ref{fig:bullet_time} and the complete video can be found in the supplementary material.
Note that we do not synthesize background, i.e., the rotating floor and walls, but render them with Blender\footnote{https://www.blender.org/} with the same orbiting camera.
Then, we segment out the foreground of our synthesized video, using the method of~\cite{Cae+17}, and composite the foreground and the rendered background.

\subsection{User Study}
Following many other image synthesis methods, we evaluate our approach in terms of user perception via a user study and also provide comparisons to existing methods in this manner.
Therefore, we show pairs of video synthesis results from 6 different methods to 18 users.
These six methods include ours and the methods of Esser et al. \cite{Esser2018}, Ma et al. \cite{MaSGVSF2017}, Liu et al. \cite{Liu2018Neural}, Chan et al. \cite{Chan2018}, and Wang et al. \cite{wang2018vid2vid}.
Our result is always included in each pair, thus performing the direct comparison between our method and each of the existing methods.
In total, 30 pairs of videos from two sequences are shown to the users.
The user study video and the labels of all pairs are provided in the supplementary material. 
After watching the videos, the users are asked to select the one from each pair that appears more natural and realistic.
In Table.~\ref{tab:user_study} we provide the percentages of votes for our method, when compared to the respective existing method.
We can see that our results are considered more realistic than all existing methods. 
Although Wang et al. \cite{wang2018vid2vid} is slightly more preferable on sequence 2, we show in the supplementary video that their method only transfers the appearance but incorrectly scales the person to match the driving actors shape.
Note that this user study does not allow relative comparison among the previous methods, since they are not directly shown to the user side by side.

\begin{table}[h]
    \centering
    \small
    \caption{Comparison of our method with existing methods through a user study. The percentages of votes for our method are provided. Numbers larger than 50 mean that our results are considered more realistic.}
    \begin{tabular}{|l|c|c|c|}
         \hline
        \textbf{Methods} & \textbf{Seq 1} & \textbf{Seq 2} & \textbf{All} \\
         \hline
         \textbf{Esser et al. \cite{Esser2018}} & 90.74 &	94.44 &	92.59 \\
         \hline
        \textbf{Ma et al. \cite{MaSGVSF2017}} & 100.00  &	96.30  &	98.15\\
         \hline
         \textbf{Liu et al. \cite{Liu2018Neural}} & 88.68  &	72.55 &	80.61 \\
         \hline
         \textbf{Chan et al. \cite{Chan2018}} & 67.92 &	68.52 &	68.22\\
         \hline
         \textbf{Wang et al. \cite{wang2018vid2vid}} & 79.63 &	46.30 &	62.96\\
         \hline
    \end{tabular}
    \label{tab:user_study}
\end{table}

\subsection{Ablation Study}
\label{ablation}

\rev{
We evaluate the importance of individual components of our approach via a quantitative ablation study.
To this end, we split one video into a training (12130 frames) and a test set (4189 frames).
We evaluate the error on the test set with respect to the ground truth.
As we are mainly interested in synthesizing the appearance of the human body, we compute the error only on the foreground region.
}

\rev{
\textbf{Relevance of TexNet.}
First, we investigate the importance of using the dynamic texture generation based on TexNet.
For this analysis, we consider the two cases where we train the RefNet based on two alternative inputs: 1) the static texture from the 3D reconstruction (cf.~Fig.~\ref{fig:pipeline} ``Static texture''), and 2) the average texture computed from the visible texels of the texture extracted from the training video (cf.~Sec.~\ref{sec:dynamictexture}).
The reconstruction error of these two and our approach are shown in Tab.~\ref{tab:ablation_quant} (``Average texture (RefNet)'', ``Static texture (RefNet)'', and ``Ours (TexNet + RefNet)'').
We can see that our full pipeline significantly outperforms these two baseline methods in terms of average per-pixel mean error and SSIM (see the supplementary video for the visual results).}

\begin{figure}%
\includegraphics[width=1\linewidth]{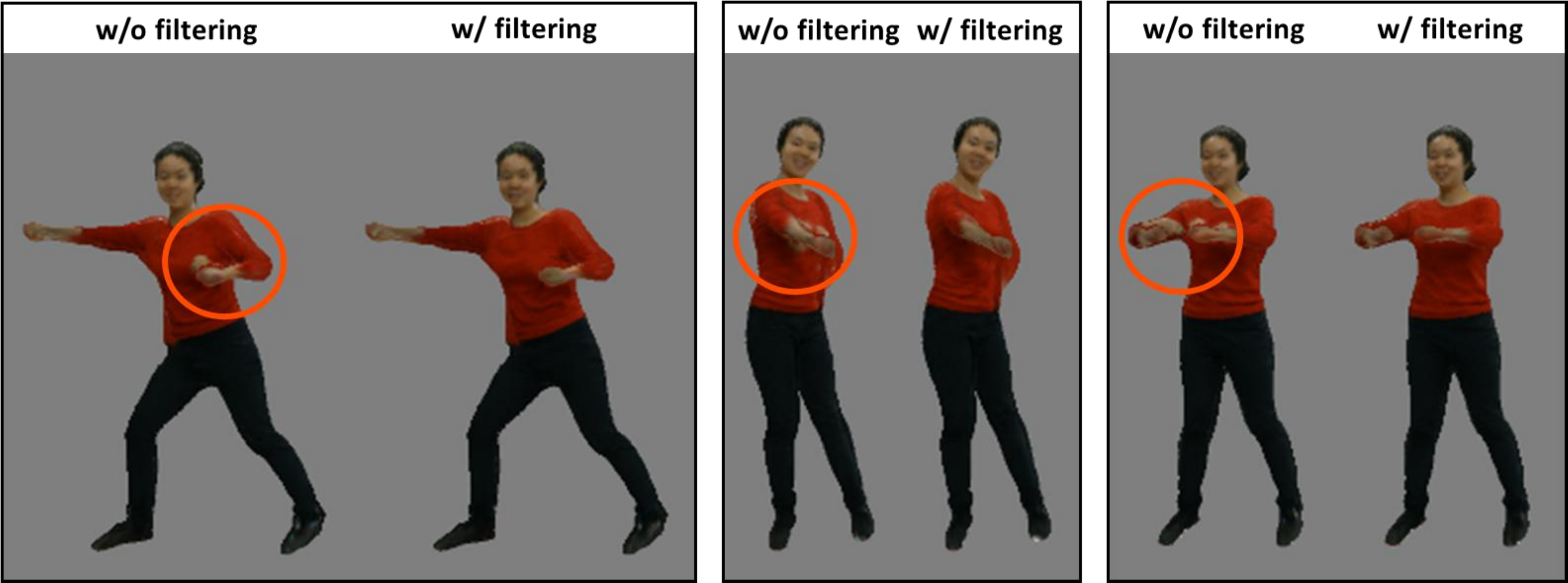}
  \caption{\rev{Ablative results for the proposed filtering procedure used for the target texture extraction. We show three instances, where the left images show the result of the rendered mesh with a dynamically generated texture without filtering, and the right images show the analogous images with filtering. When not using filtering, one can clearly see additional artifacts in the hand areas.}}
  \label{fig:ablation_filtering}
\end{figure}

\rev{
\textbf{Importance of filtering stage.}
We have also analyzed the importance of the filtering stage as used for the target texture extraction (Sec.~\ref{sec:datamodel}).
To this end, we trained one network on unfiltered data, see Tab.~\ref{tab:ablation_quant} (``Without filtering (TexNet) + RefNet'').
It can be seen that our full approach outperforms this network.
Although quantitatively the improvements may appear small due to the relatively small area that is affected, we have found that the filtering qualitatively improves the results significantly, see Fig.~\ref{fig:ablation_filtering}.
}

\rev{
\textbf{Importance of partial normal map input.}
We have also analyzed the importance of the partial normal map as input to our TexNet.
For this analysis, we consider two cases: 1) we train TexNet using a rendered 3D skeleton and its depth as input (``Rendered 3D skeleton (TexNet) + RefNet''), and 2) a direct mapping (only RefNet) from the rendered 3D skeleton to the final image (``Rendered 3D skeleton (RefNet)'').
As shown in Tab.~\ref{tab:ablation_quant}, our full pipeline outperforms these two baselines. 
\revnew{For the first case, compared to a depth map of a 3D skeleton, using a partial normal map to encode the pose as the input to TexNet is more effective and more robust since it does not need more effort to learn the translation between different domains. Also, in the second case, we can see that the dense mesh representation is more informative than the sparse skeleton and therefore can achieve better results (see the supplementary video for the visual results). }
}

\begin{table}[ht]
	\centering
	\small
	\caption{\rev{Quantitative evaluation. We report the mean (for the whole sequence) of the L2 error and SSIM for the region of the person in the foreground. Our full approach obtains the best scores.}}
	\begin{tabular}{ | l | c | c |}
		\hline
		& \textbf{L2 error} & \textbf{SSIM} \\ \hline \hline
		\textbf{\rev{Rendered 3D skeleton (TexNet) + RefNet}} & 9.558 & 0.763 \\ \hline
		\textbf{Rendered 3D skeleton (RefNet)} & 9.726 & 0.755 \\ \hline
		\textbf{Average texture (RefNet)} & 9.133 & 0.771 \\ \hline
		\textbf{Static texture (RefNet)} & 8.958 & 0.775 \\ \hline
		\textbf{Without filtering (TexNet) + RefNet} & 8.744 & 0.781 \\ \hline
		\textbf{Ours (TexNet + RefNet)} & \textbf{8.675} & \textbf{0.784} \\ \hline
	\end{tabular}
	\label{tab:ablation_quant}
\end{table}

\textbf{Size of training dataset.} We also evaluate the dependence of the performance on the size of the training dataset. In this experiment, we train TexNet and RefNet with 6000, 9000, 12130 frames of the target sequence. See Table~\ref{tab:ablation_trainsize} for the quantitative results, and the supplementary video for the visual results. As expected, larger training sets have more pose variety and hence can produce better results. For better generalizability, the poses in our training data should be as diverse as possible. If the testing pose is very different from any of the training poses, the synthesis quality will degrade but still look reasonable due to the generalizability of the networks (see, for example. the results with Youtube videos as driving sequences in the supplementary video). 

\begin{table}[ht]
	\centering
	\small
	\caption{Quantitative evaluation on the dependency of the performance on the training dataset size. We report the mean (for the whole sequence) of the L2 error and SSIM for the region of the person in the foreground. Our full training set obtains the best scores.}
	\begin{tabular}{ | l | c | c |}
		\hline
		& \textbf{L2 error} & \textbf{SSIM} \\ \hline \hline
		\textbf{6000 frames} & 10.003 & 0.749 \\ \hline
		\textbf{9000 frames} & 9.287 & 0.767 \\ \hline
		\textbf{12130 frames} & \textbf{8.675} & \textbf{0.784} \\ \hline
	\end{tabular}
	\label{tab:ablation_trainsize}
\end{table}
\section{Discussion and Limitations} 
In addition to the presented use-cases of motion transfer, interactive reenactment, and novel-view synthesis, another potential application of our approach is the generation of annotated large-scale human image or video datasets.
Particularly, with the recent popularity of deep learning, such datasets could be used for many different computer vision tasks, such as human detection, body pose estimation, and person re-identification.

Our experimental results demonstrate that our method outperforms previous approaches for the synthesis of human videos.
However, there are still some issues that could be addressed in future work.
One important issue is that the currently used neural network architectures (TexNet and RefNet) are computationally expensive to train. In order to move on to very high image resolutions, one needs to reduce the network training time. For example, training each network for an image resolution of $256 \times 256$ takes already two days, and training it for an image resolution of $512 \times 512$ takes about $6$ days on 8 high-end GPUs, and training for an image resolution of $1024 \times 1024$ takes about 10 days on 8 high-end GPUs.
Another point that is a common issue in machine learning approaches is generalization. On the one hand, our trained networks can only produce reasonable results when the training data has a similar distribution to the test data. For example, it would not be possible to train a network using frontal body views only, and then synthesize reasonable backsides of a person. On the other hand, in our current approach we train person-specific networks, whereas it would be desirable to train networks for more general settings. While we cannot claim that the results produced by our approach are entirely free of artifacts, we have demonstrated that in overall the amount and severity of artifacts is significantly reduced compared to other methods.
\rev{Another limitation is that we are not able to faithfully generate the fingers, since the human performance capture method cannot track finger motion. This can be alleviated in future works by incorporating a more complicated hand model and finger tracking components.}
Furthermore, the artifacts regarding the hands and feet are due to the 3D tracking used for generating the training data. The error in the 3D tracking would lead to a misalignment between the ground truth image and the rendered mesh in the second stage, which makes it hard for the network to directly learn this mapping.
\section{Conclusion} 
We have presented a novel method for video synthesis of human actors. 
Our method is a data-driven approach that learns, from a monocular video, to generate realistic video footage of an actor, conditioned on skeleton pose input.
In contrast to most existing methods that directly translate the sparse pose information into images, our proposed approach explicitly disentangles the learning of time-coherent fine-scale pose-dependent details from the embedding of the human in 2D screen space.
As a result, our approach leads to significant better human video synthesis results, as we have demonstrated both  qualitatively and quantitatively.

\section*{Acknowledgment}
We thank our reviewers for their invaluable comments. We also thank Liqian Ma and Caroline Chan for their great help with comparison; Neng Qian, Vladislav Golyanik, Yang He, Franziska Mueller and Ikhsanul Habibie for data acquisition; Gereon Fox for audio recording; Jiatao Gu and Daniele Panozzo for discussion. This work was supported by ERC Consolidator Grant 4DReply (770784), Lise Meitner Postdoctoral Fellowship, Max Planck Center for Visual Computing and Communications (MPC-VCC) and the Research Grant Council of Hong Kong (GRF 17210718).

\bibliographystyle{IEEEtran}
\bibliography{main_bib}

\end{document}